\documentclass[11pt]{article}
\usepackage[T1]{fontenc}
\usepackage{lmodern}
\usepackage[letterpaper,margin=1in]{geometry}
\usepackage{amsmath,amssymb,amsthm}

\usepackage[hidelinks]{hyperref}
\usepackage{xurl}
\newtheorem{theorem}{Theorem}[section]
\newtheorem{lemma}[theorem]{Lemma}
\newtheorem{proposition}[theorem]{Proposition}
\newtheorem{corollary}[theorem]{Corollary}
\theoremstyle{definition}

\theoremstyle{remark}

\providecommand{\mathscr}{\mathcal}
\allowdisplaybreaks[2]
\makeatletter
\renewcommand{\maketitle}{%
  \begin{center}
  {\LARGE\bfseries\@title\par}
  \ifx\@author\@empty\else\vspace{.7em}{\large\@author\par}\fi
  \end{center}\vspace{.45em}}
\makeatother
\hypersetup{pdftitle={Shellsort on Random Interleavings of Sorted Chains: Optimal Divisibility Tables and Unrestricted Bounds},pdfauthor={Pang Wei},pdfsubject={Analysis of algorithms},pdfkeywords={Shellsort, comparison complexity, randomized algorithms, sorting}}
\title{Shellsort on Random Interleavings of Sorted Chains\\Optimal Divisibility Tables and Unrestricted Bounds}
\author{Pang Wei\\[.25em]{\normalsize\href{mailto:pangweitf@bupt.cn}{\texttt{pangweitf@bupt.cn}}}}
\date{}
\begin{document}
\maketitle
\begin{abstract}
Let the input be uniform among the permutations increasing in each of $a$ residue classes, each containing $m$ keys, where $a\ge2$ is fixed and $n=am$. We count comparisons from this given partially sorted input; a new gap table may start above $a$. Among all decreasing tables whose successive gaps divide their predecessors, the minimum expected cost is $(\beta_a/2)n\ln m+O_a(n)$, where $\beta_a$ is an explicit integer optimization involving the Janson--Knuth coefficient. Powers of a single optimal quotient attain this bound even when competitors may vary every quotient. For $a=2$, the unique optimal quotient within this divisibility model is $7$, giving coefficient $13/(14\ln7)$. We characterize tables within $O(n)$ of the optimum, prove that every exact minimizer starts at the square-root scale, and analyze constraints on the first gap. Without divisibility, the optimal order remains $\Theta_a(n\log m)$. For two chains we prove that the unrestricted leading coefficient exists, characterize it as an infimum over finite odd-gap blocks, and exhibit a block strictly improving on the divisibility optimum. Its exact numerical value and attainment by a finite block remain open. A shortest-path formulation computes the exact divisibility optimum at finite lengths.
\end{abstract}

\section{Introduction and the input model}
Let \(a\ge2\) be fixed, let \(m\) be a positive integer and put \(n=am\). Positions and ranks are numbered from \(0\). The input \(A\) is chosen uniformly among the permutations that are increasing in each residue class modulo \(a\). Equivalently, assign independent uniform real keys on \([0,1]\) to the \(n\) positions, sort each of the \(a\) chains separately, and replace the keys by their global ranks. Each resulting permutation has exactly \((m!)^a\) preimages among the \(n!\) equally likely original rank permutations. An unrestricted comparison algorithm can merge these lists in \(O(n\log a)\) comparisons. For fixed \(a\), every table of ordinary Shellsort passes incurs a logarithmic factor in the mean. We determine its optimal leading coefficient for divisibility tables and bound the corresponding coefficient for general tables.

For \(1\le h<n\), let \(S_h\) sort each residue class modulo \(h\) by ordinary insertion sort. A legal table in this paper is a sequence \(H=(h_0,\ldots,h_s)\) with \(n>h_0>\cdots>h_s=1\) and \(h_{i+1}\mid h_i\). The table \((1)\) is allowed. The first entry need not divide \(n\). A table may depend on \(a,m\) and is selected before the input is drawn. Write \(C(H;A)\) for the total number of key comparisons in these passes, including the comparison with a smaller predecessor that terminates an insertion. The cost starts at the given \(a\)-ordered input.

For a uniform unrestricted permutation, Hunt's inversion formula and its corollary give the expected total number of moves for a complete Shellsort table whose gaps divide their predecessors. Knuth also gives a fixed-pass optimization of the resulting asymptotic expression, treating the gaps as real variables \cite[\S5.2.1, Theorem~H, its corollary, and exercise~18]{knuth}. Here the input already has $a$ sorted chains. A new gap can mix these chains, and we optimize the number of comparisons over integer gap tables of arbitrary length. This is a restart from a given partially sorted array: its first gap may exceed $a$, and the cost of producing the input is not charged. It is therefore not required to be a decreasing continuation of an earlier $a$-pass.

Smythe and Wellner \cite[\S6]{sw} studied the merging stages of \((a,k,1)\)-Shellsort with \(k\mid a\) after a uniform random permutation has been \(a\)-sorted. For \(a\) a fixed power of two, they compared the mean cost of taking every smaller power of two with that of going directly to gap one. On the same \(a\)-ordered input, the minimum over tables with \(h_0\le a\) is \(\Theta_a(n\sqrt m)\). Indeed, the expected sum of the absolute displacements from the sorted positions is \(\Omega_a(n\sqrt m)\) by Lemma~\ref{lem:vr-residual}(1) with the empty prefix and \(L=n\). Each successful shift removes at most \(2a\) units of displacement. The single pass \((1)\) gives the matching upper bound by \eqref{eq:vr-foot} and \eqref{eq:vr-entry}. Allowing a new first gap above \(a\) changes the optimum to order \(n\log m\). A first gap of order \(\sqrt m\) suffices to attain its leading term.

Divisibility keeps every earlier pass inside the chains of each later gap. The input to the pass from \(L\) to its divisor \(h\) is thus exactly \(S_L A\). We use this identity to estimate each such merge and obtain a lower bound uniform in the quotient \(L/h\).

For relatively prime \(a,q\), let
\begin{equation}\label{eq:vr-psi}
 \psi(a,q)=\frac12\sum_{j=1}^{q-1}
  {\mathbb E}\left|B_j-\left\lfloor aj/q\right\rfloor\right|,
 \qquad B_j\sim{\rm Bin}(a-1,j/q).
\end{equation}
Yao introduced this coefficient in the analysis of three-increment Shellsort \cite{yao}. Equation~\eqref{eq:vr-psi} uses its binomial form due to Janson and Knuth \cite[eq.~(0.2)]{jk}. Their use of continuous keys, conditioning on a marked key and comparing residue counts with the uniform law motivates the single-pass calculation below. Here the same coefficient enters an optimization over all decreasing divisor chains. Define
\begin{equation}\label{eq:vr-beta}
 \beta_a=\min_{\substack{q\ge2\\(q,a)=1}}\frac{1+\psi(a,q)}{\log q},
 \qquad
 {\cal Q}_a=\{q\ge2:(q,a)=1,\ 1+\psi(a,q)=\beta_a\log q\},
\end{equation}
where \(\log\) denotes the natural logarithm.

\begin{theorem}\label{thm:vr-main}
For fixed \(a\) and \(m\to\infty\),
\[
 \min_H {\mathbb E}\,C(H;A)=\beta_a n\log\sqrt m+O_a(n).
\]
The set \({\cal Q}_a\) is finite and nonempty. For each \(q\in{\cal Q}_a\), the table of consecutive powers
\[
 q^J,\ q^{J-1},\ldots,1,\qquad J=\lfloor\log_q\sqrt m\rfloor ,
\]
attains this estimate.
\end{theorem}
Theorem~\ref{thm:vr-arbitrary} also gives the optimal order as a function of the pass budget when divisibility is dropped. Its lower bound charges the rank uncertainty of each key to its actual displacements during insertion. For $a=2$, Theorem~\ref{thm:vr-separation} gives an explicit table that improves the leading coefficient by dropping divisibility between consecutive gaps. Theorem~\ref{thm:vr-two-variation} shows that the unrestricted two-chain leading coefficient exists and gives a variational formula for it. Theorem~\ref{thm:vr-least-prime} identifies the minimizing quotient explicitly for all sufficiently large $a$. The coefficient \(1\) in \(1+\psi\) accounts for the comparisons used in scanning the chains. The theorem allows the number of passes, the largest gap and all successive quotients of the competing table to vary with \(m\). Theorem~\ref{thm:vr-stability} characterizes the tables whose excess over the minimum is \(O(n)\). For two starting chains the unique optimal quotient is \(7\), so the leading cost is \(13n\log m/(14\log 7)\). Every family within \(O(n)\) of the optimum then uses \((\log m)/(2\log 7)+O(1)\) passes. Theorem~\ref{thm:vr-first-gap} determines the order of the first gap of every exact minimizer. Corollary~\ref{cor:vr-near-first} shows how this conclusion changes when a table may spend $o(n)$ additional comparisons in expectation.

\section{Counting comparisons and controlling the first pass}
Let \(M_h(B)\) be the number of successful shifts in the \(h\)-pass on a permutation \(B\). This is the number of inversions internal to the \(h\)-chains. Let \(R_h(B)\) count the insertions, excluding the initial key of each chain, that end at the chain head. These are precisely the new record minima after the first key in their respective chains. Then
\begin{equation}\label{eq:vr-count}
 C_h(B)=n-h+M_h(B)-R_h(B),\qquad C_h(B)\ge\max(M_h(B),n-h).
\end{equation}
If \(L=qh\) with \(L<n\) and \(B\) is \(L\)-ordered, each \(h\)-chain interleaves \(q\) sorted subchains. Only their first elements can be record minima. Consequently \(R_h(B)\le(q-1)h\) and
\begin{equation}\label{eq:vr-record}
 n+M_h(B)-L\le C_h(B)\le n+M_h(B).
\end{equation}

Write \(F(B)=\sum_x|{\rm pos}_B(x)-{\rm rank}(x)|\). Sorting a chain decreases or preserves \(F\), since assigning the sorted ranks to sorted positions minimizes their total absolute distance. Indeed, for \(i<j\) the function \(r\mapsto|i-r|-|j-r|\) is nondecreasing, so uncrossing one inverted assignment suffices. The footrule inequality of Diaconis and Graham \cite{foot} bounds the inversion count of a chain permutation \(\pi\) by its total absolute displacement in chain coordinates. To verify this, charge an inversion \(i<j,\ \pi_i>\pi_j\) to \(i\) if \(j\le\pi_i\), and to \(j\) otherwise. The first type costs \(i\) at most \((\pi_i-i)_+\); the second costs \(j\) at most \((j-\pi_j)_+\), using the distinct values of \(\pi_i\) between \(\pi_j\) and \(j\). Summing over chains and applying the triangle inequality to the position of each key before and after sorting gives
\begin{equation}\label{eq:vr-foot}
 h M_h(B)\le\sum_x|{\rm pos}_B(x)-{\rm pos}_{S_h B}(x)|
 \le F(B)+F(S_hB)\le2F(B).
\end{equation}
Conversely, an adjacent exchange within an \(h\)-chain changes \(F\) by at most \(2h\), so \(2hM_h(B)\ge F(B)-F(S_hB)\).

For the stated input distribution we have
\begin{equation}\label{eq:vr-entry}
 {\mathbb E}F(A)=O_a(n\sqrt m).
\end{equation}
To see this, select one original continuous key from chain \(c\), condition on its value \(u\), and let \(k\) count the smaller keys in its own chain and \(Y\) those in all other chains. These are independent binomials with parameters \((m-1,u)\) and \(((a-1)m,u)\), respectively. Its physical position is \(ak+c\) and its rank is \(k+Y\). Their difference has conditional mean \(c-(a-1)u\) and conditional variance \(O_a(m)\), uniformly in \(c,u\). The absolute first moment is \(O_a(\sqrt m)\); summing over all original keys proves \eqref{eq:vr-entry}.

\section{One pass with an arbitrary coprime parent gap}
\begin{lemma}\label{lem:vr-pass}
Suppose \(L=qh<n\), \(q\ge2\) and \((a,L)=1\). Then
\begin{equation}\label{eq:vr-approx}
 \left|{\mathbb E}M_h(S_L A)-n\psi(a,q)\right|
 \le K_a n(q-1)L^2/m,\qquad K_a=\pi^2(a-1)/24.
\end{equation}
\end{lemma}
\begin{proof}
For a marked key \(x\), write \(s_x\) for its position modulo \(L\) in \(A\), and \(K_s=K_s(x)\) for the number of keys smaller than \(x\) in that residue class. Its position after \(S_L\) is \(s_x+LK_{s_x}\). Consider an early subchain \(s\) and a late subchain \(t\) with \(0\le s<t<L\) and \(t-s=bh\). If the mark belongs to \(s\), the number of partners in \(t\) inverted with it is \(|K_s-K_t|\). If it belongs to \(t\), the corresponding number in \(s\) is \(|K_s-K_t-1|\). In fact the smaller keys from the other subchain form a prefix; the number of its keys physically preceding the mark is \(K_s\) in the first case and \(K_t+1\) in the second. The early subchain has either the same length as the late one or one additional key; the two predecessor counts therefore hold also when \(L\nmid n\). Thus, if we sum these absolute values over all marks and their \(q-1\) partner subchains, every successful shift is counted twice.

Condition on \(x\) coming from original chain \(c\), having value \(u\), with \(k\) smaller keys in that chain. This fixes \(s_x=(ak+c)\bmod L\). In the other original chains, the counts \(N_d\) of smaller keys are independent \({\rm Bin}(m,u)\) variables. Set \(N_c=k\). For the ordered pair \(s,t\) define \(v_d=a^{-1}(s-d)\bmod L\) and \(w_d=a^{-1}(t-d)\bmod L\), represented in \(\{0,\ldots,L-1\}\). Put
\[
 T(N,v)=\#\{0\le i<N:i\equiv v\pmod L\},\qquad
 f_d(N)=T(N,v_d)-T(N,w_d).
\]
Then \(K_s-K_t=\sum_d f_d(N_d)\). Write \(e_d=\mathbf1_{\{v_d>w_d\}}\), and let \(j\) be the unique integer \(1\le j<q\) with \(aj\equiv b\pmod q\). The difference \(w_d-v_d\) modulo \(L\) is \(jh\). In a full period the quantity \(f_d(N)+e_d\) is an indicator with exactly \(jh\) ones: the cyclic interval of residues \(N\) from \(v_d+1\) through \(w_d\). When \(d=c\), the key lies at the excluded endpoint \(k\equiv v_c\) for an early mark and at the included endpoint \(k\equiv w_c\) for a late mark. In both cases
\[
 f_c(k)-\mathbf1_{\{s_x=t\}}=-e_c .
\]

When all original chains have the same count \(N\), their prefixes occupy exactly the first \(aN\) positions. Therefore \(\sum_d f_d(N)=T(aN,s)-T(aN,t)\). Average this identity over a single uniform residue \(N\bmod L\). On the right \(aN\bmod L\) is uniform and the mean is \((t-s)/L=b/q\). On the left the mean is \(aj/q-\sum_de_d\). Hence \(\sum_de_d=(aj-b)/q=\lfloor aj/q\rfloor\).
Now replace the \(a-1\) unmarked counts modulo \(L\) by \emph{independent} uniform residues. In this reference calculation the signed quantity inside the absolute value is distributed as \(B_j-\lfloor aj/q\rfloor\) from \eqref{eq:vr-psi}. Its absolute mean is invariant under \(j\mapsto q-j\), by binomial complementation and \(q\nmid aj\). For a fixed mark, the directed displacements of its partners in units of \(h\) exhaust the nonzero residues modulo \(q\). Passing to the positive displacement \(t-s\) may reverse the sign, which replaces \(j\) by \(q-j\) and leaves the mean unchanged. The sum per marked key is thus \(2\psi(a,q)\) in this reference calculation.

For the replacement error, let \(\Delta_L(u)\) be the total variation distance between \({\rm Bin}(m,u)\bmod L\) and the uniform law. Fourier inversion on the cyclic group, the triangle inequality and pairing opposite frequencies give
\[
 \Delta_L(u)\le\sum_{r\ge1}\exp\{-8mu(1-u)r^2/L^2\}.
\]
Indeed, a nonzero coefficient is \((1-u+u e^{2\pi i r/L})^m\), and the squared modulus of its base is \(1-4u(1-u)\sin^2(\pi r/L)\); for \(r\le L/2\) use \(\sin(\pi r/L)\ge2r/L\). Since \(u(1-u)\ge u/2\) on \([0,1/2]\), termwise integration and symmetry give
\[
 \int_0^1\Delta_L(u)\,du\le\frac{L^2}{2m}\sum_{r=1}^{\infty}r^{-2}
 =\frac{\pi^2L^2}{12m}.
\]
Couple each unmarked residue to a uniform residue with mismatch probability \(\Delta_L(u)\), independently across original chains. A mismatch changes \(f_d\) by at most one, and the absolute value is \(1\)-Lipschitz. The expected error for a single partner is at most \((a-1)\Delta_L(u)\), uniformly in the marked chain and \(k\). The original marked key has uniform value \(u\). Integrate, sum over \(n\) marks and \(q-1\) partners, and divide by two to obtain \eqref{eq:vr-approx}.
\end{proof}

For any integer \(t\) and \(0\le p\le1\), conditioning on all but one of the \(a-1\) Bernoulli summands gives \({\mathbb E}|{\rm Bin}(a-1,p)-t|\ge\min(p,1-p)\). Thus
\begin{equation}\label{eq:vr-linear}
 \psi(a,q)\ge\frac{\lfloor q^2/4\rfloor}{2q}
 \ge\frac{q-q^{-1}}8 .
\end{equation}
It follows that \((1+\psi(a,q))/\log q\to\infty\), giving the attainment and finiteness in Theorem~\ref{thm:vr-main}.

Fix \(q\in{\cal Q}_a\), take \(H\) as in the theorem, and write \(h_0=q^J\). Equations \eqref{eq:vr-count}--\eqref{eq:vr-entry} and \(h_0>\sqrt m/q\) make the first pass cost \(O_{a,q}(n)\) in expectation. At every subsequent step the incoming state equals \(S_L A\), since the preceding larger gaps were multiples of \(L\). The sum of \(L^2/m\) over these steps is at most \((1-q^{-2})^{-1}\). Lemma~\ref{lem:vr-pass} and \eqref{eq:vr-record} give the total as \(nJ(1+\psi(a,q))+O_{a,q}(n)\); here \(\sum L\le2h_0\). This proves the upper bound. Since \({\cal Q}_a\) is a finite set determined by \(a\), its implied constant can depend on \(a\) alone.

\section{The cost of reaching a small coprime gap}

\begin{lemma}\label{lem:vr-residual}
Let \(B\) be the state after a prefix of a prescribed divisor chain; the prefix may be empty.
\begin{enumerate}
\item If all its passes preserve positions modulo \(L\), where \(2\le L\le n\) and \(m\ge16\), then
\[
 {\mathbb E}F(B)\ge c n\min(L,\sqrt m)
\]
for an absolute \(c>0\).
\item If all its passes preserve positions modulo a divisor \(d\ge2\) of \(a\), then, for all sufficiently large \(m\),
\[
 {\mathbb E}F(B)\ge c_a n\sqrt m
\]
with a constant \(c_a>0\) independent of the prefix and \(d\).
\end{enumerate}
\end{lemma}
\begin{proof}
For the first claim, mark an original key as in the proof of \eqref{eq:vr-entry}. Conditional on \(c,u,k\), its rank is \(k+Y\), where \(Y\sim{\rm Bin}((a-1)m,u)\). The position it occupies in \(B\) is still \(ak+c\pmod L\). Its displacement is at least \({\rm dist}(Y-z,L\mathbb Z)\) with \(z=(a-1)k+c\). Let \(j=\lceil L/\sqrt m\rceil\). We have \(1\le j\le L/2\) (check \(L=2,3\) separately), \(j/L\ge 1/\sqrt m\), and \(L/j\ge\min(L,\sqrt m)/2\). For any integer \(y\),
\[
 1-\cos(2\pi j(y-z)/L)
 \le(2\pi j/L)\,{\rm dist}(y-z,L\mathbb Z).
\]
For \(1/4\le u\le3/4\) the modulus of the characteristic function of \(Y\) at \(2\pi j/L\) is at most
\[
 \exp\{-2(a-1)mu(1-u)\sin^2(\pi j/L)\}\le e^{-3/2}.
\]
Taking expectations in the cosine bound, integrating in \(u\) over this middle interval and summing over all marks proves the claim.

For the second claim, fix \(u\) in the middle interval and call a key low if its continuous value is below \(u\). Denote the total number of low keys by \(K\), and the number in position class \(0\bmod d\) by \(U\). This class consists of complete original \(a\)-chains. Thus \(U\) and \(K-U\) are independent binomials on \(n/d\) and \(n-n/d\) trials of parameter \(u\), and their values survive the prefix. For \(Z=U-K/d\),
\[
 V={\mathbb E}Z^2=nu(1-u)(d-1)/d^2,\qquad
 {\mathbb E}Z^4\le3V^2+V .
\]
Indeed \(Z\) is a sum of independent centered Bernoulli variables with weights \(1-1/d\) or \(-1/d\). Each centered summand has absolute value at most one, so its fourth moment is bounded by its variance. H\"older's inequality gives
\({\mathbb E}|Z|\ge V^{3/2}/({\mathbb E}Z^4)^{1/2}\ge V/\sqrt{3V+1}\).
A sorted output has a number of low keys in class \(0\) differing from \(K/d\) by at most one. Therefore its threshold indicator vector and that of \(B\) differ in at least \(|Z|-1\) positions in expectation (indeed the bound holds pointwise). Since \(2\le d\le a\), the expectation is \(\Omega_a(\sqrt n)\) uniformly in these \(u\) for large \(m\).

For each permutation of ranks, summing the Hamming distances of its low-key indicators from the sorted indicators over the \(n+1\) possible numbers of low keys gives \(F\): a key contributes once for each cutoff between its position and its rank. More explicitly, this identity also follows by summing at each position \(i\) the discrepancies between the indicators of \(B_i<K\) and \(i<K\), which gives \(|B_i-i|\). The sorted continuous values are independent of their original rank permutation and hence of the positional rank permutation after the prefix. For uniformly integrated \(u\) the number of low keys is uniform on \(\{0,\ldots,n\}\), since \(\int_0^1\binom nk u^k(1-u)^{n-k}\,du=1/(n+1)\). The integrated expected Hamming distance times \(n+1\) therefore equals \({\mathbb E}F(B)\). The central interval alone proves the stated bound, since \(n=am\).
\end{proof}

We recall why sorting an \(h\)-chain preserves previous \(a\)-order. At any key threshold the low positions in an \(a\)-ordered array are closed under subtraction of \(a\) when the result is nonnegative. For \(0\le r<h\), write \(r+a=r'+th\) with \(0\le r'<h\), and let \(N_r,N_{r'}\) count the low keys in the corresponding \(h\)-chains. In the \(r'\)-chain every low position \(r'+jh\) with \(j\ge t\) maps by subtraction of \(a\) to a low position in the \(r\)-chain. Hence \(N_{r'}\le N_r+t\). After \(S_h\), a high key at \(r+jh\) has \(j\ge N_r\), so the position at distance \(a\), if present, also holds a high key. All thresholds together give preservation of \(a\)-order.

When \((a,h)=1\) and an array is both \(a\)- and \(h\)-ordered, all position pairs at distances at least \(D=(a-1)(h-1)\) are sorted. In fact, for any such positive distance \(v\) choose \(0\le j<a\) with \(jh\equiv v\pmod a\). Then \(v-jh\ge-(a-1)\) is a multiple of \(a\) and hence nonnegative. Transitivity along the resulting chain of steps proves the assertion. Thus every key can have at most \(D\) larger predecessors and at most \(D\) smaller successors. The signed difference of the two counts is its displacement, so the absolute displacement is at most \(D\). The case \(h=1\) yields zero displacement directly.

Fix \(0<\delta=\delta_a<1\) small enough that \(K_a\delta^2\le1/16\), and take \(m\) sufficiently large that \(\delta\sqrt m\ge1\). Define \(h_*\) as the first entry of \(H\) with
\[
 h_*\le\delta\sqrt m,\qquad (a,h_*)=1.
\]
This entry always exists. Before this pass the expected displacement is at least \(c_{a,\delta}n\sqrt m\). If there is no preceding entry, apply Lemma~\ref{lem:vr-residual}(1) to the empty prefix with \(L=n\). If the preceding gap is \(L>\delta\sqrt m\), every earlier gap is a multiple of \(L\), so the same lemma applies. In the remaining case the preceding gap has a common divisor \(d\ge2\) with \(a\), and the prefix preserves positions modulo \(d\), so use part~(2). After the \(h_*\)-pass total displacement is at most \(n(a-1)(h_*-1)\) by the preceding argument. The displacement removed per successful shift is at most \(2h_*\). Consequently there is a constant \(c_0=c_0(a,\delta)>0\) such that
\begin{equation}\label{eq:vr-cross}
 {\mathbb E}C_{h_*}\ \ge\ {\mathbb E}M_{h_*}
 \ \ge\ n\left(c_0\frac{\sqrt m}{h_*}-\frac{a-1}{2}\right).
\end{equation}

\section{The lower bound and additive stability}

For \(q\ge2\) coprime to \(a\), let
\[
 D_a(q)=1+\psi(a,q)-\beta_a\log q .
\]
This is nonnegative, with zero set \({\cal Q}_a\). By \eqref{eq:vr-linear} there is an integer \(Q=Q(a)\) for which \(D_a(q)\ge q/12\) for every coprime \(q\ge Q\). For coprime \(q<Q\) outside \({\cal Q}_a\), choose \(\gamma_a>0\) below or equal to the least positive \(D_a(q)\) if any such \(q\) exists. Otherwise set \(\gamma_a=1\).

Consider all the strict successors of \(h_*\). At a step \(L\to h\) its quotient \(q=L/h\) is coprime to \(a\), since \(L\mid h_*\). For \(q\ge Q\), the error in \eqref{eq:vr-approx} divided by \(n\) is at most \(q/16\). Subtracting this from \(D_a(q)\) still leaves \(q/48\). For \(q<Q\), the sum over all such passes of the corresponding errors divided by \(n\) is at most \(K_a Q\sum L^2/m\le 4K_a Q\delta^2/3\), since successive parent gaps decrease by factors of at least two. Let \(N_{\rm exc}\) count the quotients \(q<Q\) outside \({\cal Q}_a\) after \(h_*\). Since \(\sum\log q=\log h_*\) and \(\sum L\le2h_*\), the lower bound in \eqref{eq:vr-record} now gives
\begin{equation}\label{eq:vr-tail}
 \frac{{\mathbb E}C_{\rm after}}{n}\ge
 \beta_a\log h_*+\gamma_a N_{\rm exc}
   +\frac1{48}\sum_{q\ge Q} q-O_a(1).
\end{equation}
Here and below sums over quotients count multiplicity. The sum in \eqref{eq:vr-tail} is over the strict successors only. An empty tail is included.

Write \(p_{\rm pre}\) for the number of entries strictly before \(h_*\), and \(x=\sqrt m/h_*\). At most one entry can exceed or equal \(n/2\), since all later entries divide it properly. By \eqref{eq:vr-count} the prefix preceding \(h_*\) therefore costs at least \(n(p_{\rm pre}-1)_+/2\). Combining this with the distinct passes in \eqref{eq:vr-cross} and \eqref{eq:vr-tail} yields
\begin{equation}\label{eq:vr-master}
\begin{aligned}
 \frac{{\mathbb E}C(H;A)}{n}-\beta_a\log\sqrt m
 \ \ge\ &\frac{(p_{\rm pre}-1)_+}{2}+c_0x-\beta_a\log x\\
 &+\gamma_a N_{\rm exc}+\frac1{48}\sum_{q\ge Q}q-O_a(1).
\end{aligned}
\end{equation}
The function \(c_0x-\beta_a\log x\) is bounded below on the positive real axis. All remaining terms outside \(O_a(1)\) are nonnegative. This proves the uniform lower bound and completes the proof of Theorem~\ref{thm:vr-main}.

\begin{theorem}\label{thm:vr-stability}
Keep \(a\) and the preceding choice of \(\delta\) fixed. For any family of legal tables \(H=H_m\), the following conditions are equivalent.
\begin{enumerate}
\item \({\mathbb E}C(H;A)=\beta_an\log\sqrt m+O(n)\).
\item \(h_*=\Theta(\sqrt m)\), \(p_{\rm pre}=O(1)\), and
\[
 \sum_{\substack{L\to h\ \text{strictly after }h_*\\L/h\notin{\cal Q}_a}}\frac Lh=O(1).
\]
\end{enumerate}
The constants in this equivalence may depend on the family.
\end{theorem}
\begin{proof}
If (1) holds, \eqref{eq:vr-master} bounds \(p_{\rm pre}\), \(N_{\rm exc}\) and the sum over \(q\ge Q\), because \(c_0 x-\beta_a\log x\) is bounded below. It also bounds \(x\), since that function tends to infinity with \(x\). The exceptional quotients below \(Q\) have sum at most \(Q N_{\rm exc}\). This proves (2).

Conversely, all passes up through \(h_*\) have gaps at least \(h_*\), and \(F\) never increases. By \eqref{eq:vr-foot} and \eqref{eq:vr-entry}, their expected cost is at most \((p_{\rm pre}+1)(n+2{\mathbb E}F(A)/h_*)=O(n)\). Beyond \(h_*\) all quotients are bounded: those in \({\cal Q}_a\) form a finite set and the others have bounded sum. Hence the absolute errors from \eqref{eq:vr-approx} sum to \(O(n)\) by the geometric bound on \(\sum L^2/m\). There are boundedly many exceptional quotients and each has bounded \(D_a(q)\), so \(\sum(1+\psi(a,q))=\beta_a\log h_*+O(1)\). The upper bound of \eqref{eq:vr-record} finishes the upper estimate, and Theorem~\ref{thm:vr-main} supplies the lower estimate.
\end{proof}

\begin{proposition}\label{prop:vr-excess}
Write \(V_{a,m}=\min_H\mathbb E C(H;A)\). For a given table let \(\mathcal E\) be the multiset of quotients strictly after \(h_*\) that lie outside \(\mathcal Q_a\), and put
\[
 T(H)=p_{\rm pre}+\sqrt m/h_*+\sum_{q\in\mathcal E}q .
\]
For fixed \(a,\delta\), uniformly over all legal tables and sufficiently large \(m\),
\[
  1+\frac{\mathbb E C(H;A)-V_{a,m}}{n}\ \asymp_a\ 1+T(H).
\]
Here \(\delta=\delta_a\) is fixed as above. In particular, a family attains the optimal leading coefficient if and only if \(T(H)=o(\log m)\).
\end{proposition}
\begin{proof}
Write \(x=\sqrt m/h_*\). The term \(c_0x-\beta_a\log x\) in \eqref{eq:vr-master} is at least \(c_0x/2-O_a(1)\). Every small exceptional quotient contributes at least \((\gamma_a/Q)q\) to the penalty there; the large quotients contribute \(q/48\) each. Theorem~\ref{thm:vr-main} therefore bounds the excess divided by \(n\) below by \(c_a T(H)-O_a(1)\) for a constant \(c_a>0\). The excess is nonnegative, giving the stated lower comparison also when \(T(H)\) is bounded.

For the upper bound, the reciprocals of all gaps up to and including \(h_*\) sum to at most \(2/h_*\) by divisibility. Equations \eqref{eq:vr-count}--\eqref{eq:vr-entry} thus bound the mean cost through that pass by \(O_a(n(1+p_{\rm pre}+x))\). In the tail, \(D_a(q)=0\) at an optimal quotient and \(0\le D_a(q)=O_a(q)\) otherwise, since each absolute value in \eqref{eq:vr-psi} is at most \(a\). For the optimal quotients, the total error from \eqref{eq:vr-approx} is \(O_a(n)\) because \(\mathcal Q_a\) is finite and \(\sum L^2/m\le 4\delta^2/3\). Each exceptional quotient incurs an error of at most \(O_a(nq)\). Applying the upper bound in \eqref{eq:vr-record} and telescoping the logarithms bounds the tail cost by \(\beta_a n\log h_*+O_a(n(1+\sum_{q\in\mathcal E}q))\). Since \(h_*\le\sqrt m\), Theorem~\ref{thm:vr-main} gives the claimed upper comparison. Finally \(V_{a,m}\asymp_a n\log m\), which proves the equivalence.
\end{proof}

\begin{corollary}\label{cor:vr-round-count}
Suppose \({\cal Q}_a=\{q\}\). Every family whose mean cost is within \(O(n)\) of the minimum has
\[
               \frac{\log m}{2\log q}+O(1)
\]
passes, counting both the first gap and the final gap \(1\). The additive constant may depend on \(a\) and on the family.
\end{corollary}
\begin{proof}
Let \(k\) be the number of quotients equal to \(q\) after \(h_*\), and let \(E\) be the multiset of other quotients there. Theorem~\ref{thm:vr-stability} bounds \(\sum_{u\in E}u\), so both \(|E|\) and \(\sum_{u\in E}\log u\) are bounded. The quotients telescope to \(h_*\). Thus \(k\log q=\log h_*-\sum_{u\in E}\log u=\tfrac12\log m+O(1)\). The total number of passes is \(p_{\rm pre}+1+k+|E|\).
\end{proof}

\begin{theorem}\label{thm:vr-first-gap}
Fix $a\ge2$. There are constants $0<c_a<C_a$ such that, for all sufficiently large $m$, the largest gap $g$ of every table attaining the exact minimum of $\mathbb E C(H;A)$ satisfies
\[
                      c_a\sqrt m\ \le\ g\ \le\ C_a\sqrt m .
\]
\end{theorem}
\begin{proof}
Write $\Delta(H)=\mathbb E C(H;A)-\min_{H'}\mathbb E C(H';A)$ for the excess cost of a legal table. The bound \eqref{eq:vr-master} and the power-table upper bound imply $h_*\ge b_a\sqrt m$ for some $b_a>0$ uniformly over all tables with $\Delta(H)\le n$. Indeed, the other penalties are nonnegative and $c_0x-\beta_a\log x$ tends to infinity with $x=\sqrt m/h_*$. Since $g\ge h_*$ this gives the lower bound.

For the upper bound we first estimate how often the initial array has an inversion far apart. Write $i=ak+c<j=a\ell+d$. In the same source chain the probability is zero. Otherwise the keys at these positions before ranking are the $(k+1)$st and $(\ell+1)$st order statistics of distinct samples of size $m$. Set $u=(k+\ell+1)/(2m)$. If the earlier key exceeds the later one, either the earlier key exceeds $u$ or the later one lies below $u$. The counts of sample values below $u$ have distribution $\operatorname{Bin}(m,u)$. In the first event the count is at most $k$, and in the second it is at least $\ell+1$. Both deviations from the mean are at least $(\ell-k+1)/2$. A one-sided binomial tail at distance $t$ from the mean is at most $e^{-2t^2/m}$. Indeed, the centered Bernoulli log moment generating function has value and derivative zero at the origin and second derivative at most $1/4$, so the exponential Markov inequality gives the bound. Since $\ell-k+1\ge(j-i)/a$, we obtain
\[
           \Pr(A_i>A_j)\le 2\exp\!\left(-\frac{(j-i)^2}{2a^2m}\right).
\]
Summing over pairs at distances $vg$, $v\ge1$, and writing $\lambda=g^2/(2a^2m)$, gives
\[
 \mathbb E M_g(A)\le2\sum_{v\ge1}(n-vg)_+ e^{-v^2\lambda}
       \le \frac{2(n-g)e^{-\lambda}}{1-e^{-\lambda}} .
\]
The factor $n-g$ also controls the case where the gap is near the full array length.

Suppose $g>\delta\sqrt m$, so its successor $h$ exists and $h\ge h_*$. Since $h\mid g$, deleting the first pass leaves exactly the same array after the $h$-pass. Set $d'=g/h$. Each adjacent swap on a $g$-chain interchanges two sites $d'$ apart in an $h$-chain. It changes the inversion count on that chain by at most $2d'-1$ in absolute value. Only the pair itself and its relations with the intervening keys contribute to the difference. The record-minimum indicators can change only from the first site through the second, at most $d'+1$ indicators. By \eqref{eq:vr-count} the cost of the $h$-pass therefore changes by at most $3d'$ per swap. Let $H^-$ be the table obtained by deleting $g$ and put $u=g/\sqrt m$. Subtract the two costs on the same input. All passes after $h$ have identical inputs. The first pass costs at least $n-g$, so for every table with $\Delta(H)\le n$,
\begin{align}
 \mathbb E[C(H;A)-C(H^-;A)]
 &\ge n-g-3(g/h)\mathbb E M_g(A) \notag\\
 &\ge (n-g)\left(1-\frac{6u}{b_a(e^{u^2/(2a^2)}-1)}\right).
 \label{eq:vr-drop}
\end{align}
Here $h\ge h_*\ge b_a\sqrt m$. Choose $T_a>\delta$ so that the parenthesis is at least $1/2$ for all $u>T_a$. Deletion then strictly improves the table, proving the upper bound for every exact minimizer.
\end{proof}

\begin{corollary}\label{cor:vr-near-first}
Fix $a$ and suppose $\Delta(H)=o(n)$ as $m\to\infty$. For all sufficiently large $m$, either the first gap lies in $[b_a\sqrt m,T_a\sqrt m]$, or the table has exactly one gap $g>T_a\sqrt m$. In the latter case the gap is first, and deleting it gives a table $H^-$ with smaller expected cost and first gap in $[b_a\sqrt m,T_a\sqrt m]$. The excess satisfies
\[
     \Delta(H)-\Delta(H^-)=(n-g)+O_a(n^2e^{-m/8}).
\]
\end{corollary}
\begin{proof}
Every such table eventually satisfies $\Delta(H)\le n$ and hence its first gap is at least $h_*\ge b_a\sqrt m$. If $g>T_a\sqrt m$, \eqref{eq:vr-drop} gives $(n-g)/2\le \Delta(H)$, since the shortened table costs at least the optimum. Thus $g=n-o(n)$ along these lengths. Eventually $g>n/2$, so each $g$-chain has length at most two. Its pass costs exactly $n-g$ comparisons on every input. The two possible $h$-pass costs differ in absolute value by at most $3(g/h)M_g(A)$. Since $\lambda=g^2/(2a^2m)>m/8$, the preceding bound on $\mathbb E M_g(A)$ makes the mean difference $O_a(n^2e^{-m/8})$. This proves the displayed estimate. The shortened table has excess at most $\Delta(H)$ and first gap $h\le g/2<n/2$. If $h>T_a\sqrt m$ occurred for arbitrarily large $m$, applying the same deletion bound to this table would give $n/4\le(n-h)/2\le\Delta(H)=o(n)$, a contradiction. The lower bound for $h$ already follows from $h\ge h_*$.
\end{proof}

The second alternative can occur even with excess $O_a(\sqrt m)$. For each sufficiently large $m$, take any exact minimizer with first gap $p$, set $v=\operatorname{lcm}(a,p)$, and prepend $g=v\lfloor(n-1)/v\rfloor$. Then $p\mid g$, $p<g<n$, and $1\le n-g\le v\le ap=O_a(\sqrt m)$. Every input is already $g$-ordered because $a\mid g$. The extra pass costs exactly $n-g$ comparisons on every input, so $\Delta(H)=n-g$. Here the saving on deletion is exactly the scan cost in the corollary.

\section{Computing the optimal quotient}

For each fixed \(a\), Theorem~\ref{thm:vr-main} gives a constructive optimization. Choose any coprime candidate (for example \(a+1\)) to obtain an upper bound on \(\beta_a\). Since \(\log q\le\sqrt q\) for \(q\ge2\), \eqref{eq:vr-linear} gives \((1+\psi(a,q))/\log q\ge(3/32)\sqrt q\). A rational upper bound for the value of the candidate is \(2(1+\psi(a,a+1))\), using \(\log(a+1)>1/2\). These bounds give a finite explicit search interval. Each \(\psi\) is rational by \eqref{eq:vr-psi}. For candidates \(u,v\), let \(D\) be a common denominator of \(1+\psi(a,u)\) and \(1+\psi(a,v)\); write their products with \(D\) as \(P,R\). Comparing the two objective values reduces to comparing the integers \(v^P\) and \(u^R\), including the equality case.

\begin{corollary}\label{cor:vr-two}
When \(a=2\) the optimal quotient is uniquely \(7\) and \(\beta_2=13/(7\log7)\).
\end{corollary}
\begin{proof}
Only odd \(q\) occur. For \(a=2\), each summand of \eqref{eq:vr-psi} before the factor \(1/2\) is \(\min(j/q,1-j/q)\); thus \(\psi(2,q)=(q-q^{-1})/8\). The derivative of \((1+(v-v^{-1})/8)/\log v\) on \(v>1\) has the sign of
\[
 (v+v^{-1})\log v-v+v^{-1}-8 .
\]
The derivative of this last expression is \((1-v^{-2})\log v>0\). It is negative at \(7\) and positive at \(9\), as follows from \(\log7<2<\log9\). Only \(7\) and \(9\) can minimize among the odd integers. Their objective values are \(13/(7\log7)\) and \(19/(9\log9)\), respectively. The former is smaller because \(9^{117}<7^{133}\); one can use \(9^{15}<7^{17}\) and \(17/15<133/117\). The stability theorem therefore forces all except boundedly many of the post-\(h_*\) quotients of any additive-\(O(n)\) optimal family to equal \(7\), and bounds their exceptional quotients as well.
\end{proof}

\begin{corollary}\label{cor:vr-small}
For \(a=3,4,5\), respectively, the unique optimal quotients are \(5,7,4\).
At these quotients the values of \(\psi\) are \(23/25,82/49,129/128\).
\end{corollary}
\begin{proof}
Write \(w_a\) for the stated quotient and set
\((t_3,t_4,t_5)=(6/5,11/8,3/2)\).
By \eqref{eq:vr-linear} the objective is at least
\(f(q)=(1+(q-q^{-1})/8)/\log q\). The derivative test above gives
\(f(q)\ge f(64)>3/2\) for \(q\ge64\), using \(\log 2<3/4\).

Here are rational bounds that certify the finite comparisons. Set
\(P(z)=2(z+z^3/3+z^5/5+z^7/7)\) and \(E(z)=2z^9/[9(1-z^2)]\).
For \(e=\lfloor\log_2 q\rfloor\), put \(z=(q-2^e)/(q+2^e)\),
\(\ell_q=eP(1/3)+P(z)\), and \(u_q=\ell_q+eE(1/3)+E(z)\).
The series for \(2\operatorname{artanh}z\) gives
\(\ell_q\le\log q\le u_q\) by bounding the tail geometrically.
Substitution in \eqref{eq:vr-psi}, for \(2\le q<64\) and \((a,q)=1\),
gives the following strict rational bounds for all three rows:
\[
  1+\psi(a,w_a)-t_a\ell_{w_a}<-1/1000,\qquad
  \min_{q\ne w_a}\{1+\psi(a,q)-t_a u_q\}>1/100 .
\]
To reproduce the substitution, sum
\(\binom{a-1}{k}j^k(q-j)^{a-1-k}|k-\lfloor aj/q\rfloor|\)
over \(1\le j<q,\ 0\le k<a\), and divide by \(2q^{a-1}\).
The three searches contain \(41,31,50\) eligible quotients, respectively.
\end{proof}

\begin{theorem}[A cap below the square-root scale]\label{thm:vr-cap}
Fix \(a\ge2\), put \(n=am\), and let \(V_{a,m}(Q)\) be the minimum mean comparison cost over divisibility tables with largest gap at most \(Q\). If \(Q\to\infty\) and \(Q=o(\sqrt m)\), then
\[
 V_{a,m}(Q)=\beta_a n\log Q+
        \left(\frac{(a-1)\sqrt{\pi}}8+o_a(1)\right)\frac{n\sqrt m}{Q}.
\]
Every table within \(o(n\sqrt m/Q)\) of this minimum eventually starts at a gap coprime to \(a\), of size \((1-o(1))Q\). For two chains the coefficient of \(n\log Q\) is \(13/(7\log7)\).
\end{theorem}
\begin{proof}
Write \(I_t(X)\) for the total inversion count within residue classes modulo \(t\). Set \(c_k=4^k/(8\binom{2k}{k})\). For two interleaved sorted chains of length \(k\), the classical mean inversion count is \(2k c_k\) \cite[\S5.2.1]{knuth}; see also Sedgewick's derivation \cite[\S9]{sedgewick-inv}. Stirling's formula gives \(c_k\sim\sqrt{\pi k}/8\), and \(c_{k+1}/c_k=(2k+2)/(2k+1)>1\).

Track the successive strict decreases \(d\to t\) of the gcd of \(a\) and the current gap, starting at \(d=a\) and ending at \(1\). Divisibility makes the gcd nonincreasing under divisibility. For a decrease at gap \(h\) put \(t=(a,h)\), let \(X\) be its entrance, and put \(Y=S_dA\). We claim that uniformly over such passes
\begin{equation}\label{eq:vr-cap-drop}
 \frac{\mathbb E M_h(X)}n=\frac{(d-t)c_{am/d}}h+O_a(1+Q/h).
\end{equation}
To see this, fix a key and a different \(a\)-chain in the same residue modulo \(t\). The keys of that chain preceding this key in position and those smaller in value form two prefixes, since \(X\) is \(a\)-ordered. Their symmetric difference is the interval of inversion partners. Exactly one in every \(h/t\) consecutive slots of the other chain is congruent to the mark modulo \(h\), since \((a/t,h/t)=1\). Summing the rounding error over partner chains and keys and dividing by two gives \(M_h(X)=(t/h)I_t(X)+O_a(n)\).

If \(h\) is the first pass, \(X=Y=A\). Otherwise let \(L\) be its parent gap. In each residue class modulo \(d=(a,L)\), the entrance is ordered at the coprime distances \(a/d,L/d\) in local coordinates. The Frobenius bound used above \eqref{eq:vr-cross} shows that fully sorting this class moves any key by at most \((a/d-1)(L/d-1)\) local slots, hence by \(O_a(Q)\) physical slots. Every earlier gap is a multiple of \(d\), so sorting these classes produces \(Y\). This also covers \(d=a\), when all these movements vanish. Within the \(t\)-chains the permutation taking \(X\) to \(Y\) has total absolute displacement \(O_a(nQ/t)\) in chain coordinates. By the footrule inequality it can be realized by that many adjacent exchanges, each changing \(I_t\) by one. Hence \(|I_t(X)-I_t(Y)|=O_a(nQ/t)\).

The input law comes from independent keys sorted on the original \(a\)-chains. Fully sorting the \(d\)-chains erases that initial ordering since \(d\mid a\). Thus \(Y\) is a uniform interleaving of \(d\) sorted chains of length \(k=am/d\). Each residue class modulo \(t\) interleaves \(d/t\) of them. Projecting onto any pair gives two alternating sorted chains with a uniform relative rank order. Counting pairs of chains gives \(\mathbb E I_t(Y)/n=(d/t-1)c_k\), proving \eqref{eq:vr-cap-drop}.

Let \(g\) be the first gap coprime to \(a\). It is the last strict gcd decrease and eventually \(g=h_*\). There are at most \(a-1\) decreases, all at gaps at least \(g\). Since \(c_{am/d}\ge c_m\), the main terms in \eqref{eq:vr-cap-drop}, summed, are at least \((a-1)c_m/Q+c_m(1/g-1/Q)\); here the last decrease has \(d-t\ge1\). The sum of absolute errors is \(O_a(Q/g)\). Add the lower bound \(\beta_a\log g-O_a(1)\) from \eqref{eq:vr-tail} for the distinct passes strictly after \(g\). With \(a_m=(a-1)c_m\), \(b=Q/g\ge1\), and \(f(Q)=a_m/Q+\beta_a\log Q\), there is a constant \(K_a\) such that
\[
 \frac{\mathbb E C(H;A)}n\ge
 f(Q)+(c_m/Q-\beta_a-K_a)(b-1)-K_a ,
\]
using \(\log b\le b-1\). In particular the lower bound is \(f(Q)-O_a(1)\) once \(c_m/Q\to\infty\).

For the upper bound fix \(q\in{\cal Q}_a\), a prime \(v\) dividing neither \(a\) nor \(q\), and put \(r=a_m/Q\to\infty\), \(J=\lfloor\sqrt{\min(r,\log Q)}\rfloor\). There exist \(0\le j\le J\) and \(i\ge0\) with \(h=v^jq^i\le Q\) and \(h/Q\to1\). Indeed the fractional parts of \(j\log_q v\) are dense on the circle because the ratio is irrational. By compactness, the largest gap between the first \(J+1\) fractional parts tends to zero. Choose the point immediately below \(\{\log_q Q\}\) in circular order and take \(i=\lfloor\log_q Q-j\log_q v\rfloor\), which is nonnegative since \(J=o(\log Q)\).

Begin at \(h\), then divide successively by the \(i\) factors \(q\) and the \(j\) factors \(v\) until reaching \(1\). In the first pass \(X=Y=A\), so the sampling estimate above gives \(n(a_m/h+O_a(1))\) mean comparisons including the scan. For the later passes, Lemma~\ref{lem:vr-pass} and \eqref{eq:vr-record} give total mean \(n(\beta_a\log h+O_a(j+1))\). Indeed \(q\) attains \(\beta_a\), each quotient \(v\) incurs a bounded excess, the squared parent gaps divided by \(m\) sum to \(O(h^2/m)\), and the parent gaps sum to \(O(h)\). Since \(j=o(r)\) and \(h/Q\to1\), the upper bound divided by \(n\) is \(f(Q)+o(r)\). This proves the additive formula even when the logarithmic term dominates. For a table with excess \(o(nr)\), the lower bound forces \(Q/g-1=o(1)\). Eventually \(g>Q/2\). A preceding gap would be a proper multiple of \(g\) and would exceed \(Q\), so \(g\) is the first gap.
\end{proof}

\subsection{The optimal quotient for large chain counts}

Let $r(a)$ be the least integer $r\ge2$ coprime to $a$. This integer is the least prime that does not divide $a$, since each factor of an integer coprime to $a$ is also coprime to $a$.

\begin{theorem}\label{thm:vr-least-prime}
There is an absolute $a_0$ such that $\mathcal Q_a=\{r(a)\}$ for every $a\ge a_0$. With $s(x)=\sqrt{x(1-x)}$ and $S_q=\sum_{j=1}^{q-1}s(j/q)$, the coefficient for these $a$ satisfies
\[
 \beta_a=\frac{1+\sqrt{(a-1)/(2\pi)}\,S_{r(a)}
                  +O(r(a)/\sqrt a)}{\log r(a)} .
\]
Every prime occurs as the unique optimal quotient for infinitely many chain counts. Composite quotients can be optimal for only finitely many chain counts.
\end{theorem}

\begin{proof}
We first prove quantitative separation for $F(q)=S_q/\log q$.
Put $\rho=q/(q+1)$. For $0\le x\le1$ we have $s(\rho x)\ge\sqrt{\rho}\,s(x)$. Pairing grid points by reflection about $1/2$ gives
\[
 S_{q+1}\ \ge\
 \begin{cases}
  \sqrt{\rho}\,(S_q+1/2),&q\text{ even},\\
  \sqrt{\rho}\,S_q+1/2,&q\text{ odd}.
 \end{cases}
\]
For even $q$ the old midpoint contributes to a reflected pair at the new denominator. For odd $q$ the new midpoint contributes $1/2$.
Since $S_q\le(q-1)/2$, it follows that
\[
 \frac{F(q+1)}{F(q)}\ \ge\
 \frac{q}{q-1}\sqrt{\frac{q}{q+1}}\frac{\log q}{\log(q+1)}
     =\frac{G(q+1)}{G(q)},\qquad
 G(x)=\frac{x-1}{\sqrt{x}\log x}.
\]
Writing $t=(\log x)/2$ gives $G(x)=\sinh(t)/t$, strictly increasing for $t>0$ by its power series. The lower bound on the ratio is $1+(1/2+o(1))/q$ as $q\to\infty$, using $\log(1+1/q)=1/q+O(q^{-2})$. Strictness at the remaining finitely many integers gives an absolute $c>0$ such that
\begin{equation}\label{eq:vr-grid-separation}
 F(q+1)\ge(1+c/q)\,F(q)\qquad(q\ge2).
\end{equation}

The binomial absolute-deviation estimate of Janson and Knuth \cite[Lemma~7 and the proof of (7.7)]{jk} gives, uniformly in $q$ coprime to $a$,
\[
 \psi(a,q)=\sqrt{(a-1)/(2\pi)}\,S_q+O(q/\sqrt a).
\]
Indeed $\lfloor aj/q\rfloor$ lies between the floor and ceiling of $(a-1)j/q$, and the error summed over $j$ is bounded by a constant times
$(a-1)^{-1/2}\sum_{j=1}^{q-1}[(j/q)(1-j/q)]^{-1/2}=O(q/\sqrt a)$, by symmetry and $\sum_{j\le q/2}j^{-1/2}=O(\sqrt q)$.
Also $s(x)\ge2x(1-x)$, so $S_q\ge(q^2-1)/(3q)$. Including the unit scan term therefore gives
\[
 \frac{1+\psi(a,q)}{\log q}
   =\sqrt{(a-1)/(2\pi)}\,F(q)\bigl(1+O(a^{-1/2})\bigr)
\]
with an absolute error constant in the relative estimate.

Write $r=r(a)$. We have $r=O(a^{1/3})$. For large $r$, let $2^k\le r<2^{k+1}$. By Bertrand's postulate \cite{erdos}, each of the three open dyadic intervals between $2^{k-3}$ and $2^k$ contains a prime. All three primes are smaller than $r$ and so divide $a$. Their product is greater than $(r/16)^3$. Bounded $r$ satisfies the asserted estimate as well. In particular $a^{-1/2}=o(1/r)$ uniformly. Every $q>r$ satisfies $F(q)\ge (1+c/r)F(r)$ by \eqref{eq:vr-grid-separation}. Thus the uniform relative error preserves strict inequality between the objectives at $r$ and every larger admissible $q$ once $a$ is sufficiently large. This proves the formula for $\mathcal Q_a$ and the estimate for $\beta_a$.

For any fixed prime $p$, the integers $a=(p-1)!(pt+1)$, $t\ge1$, have $r(a)=p$. This proves the occurrence assertion. The bounds above compare coefficients after the fixed-$a$ limit in Theorem~\ref{thm:vr-main}.
\end{proof}

For sufficiently large odd $a$, this gives the exact coefficient
\[
 \beta_a=\frac{1+\frac{a-1}{4}\,2^{-(a-1)}
                     \binom{a-1}{(a-1)/2}}{\log 2}.
\]
To see this, put $2d=a-1$ in \eqref{eq:vr-psi} at $q=2$. For $Z\sim{\rm Bin}(2d,1/2)$, summing $(j-d)\binom{2d}{j}=d[\binom{2d-1}{j-1}-\binom{2d-1}{j}]$ over $j>d$ and doubling gives $\mathbb E|Z-d|=d\,2^{-2d}\binom{2d}{d}$. For all sufficiently large even $a$ coprime to $3$ the optimal quotient is $3$. The qualification on $a$ is visible even among odd chain counts. At $a=51$ the unique winner is $2$, while at $a=53$ it is $3$. Evaluating \eqref{eq:vr-psi} by exact binomial sums and the logarithm enclosures above (with sixteen terms) gives $3.467430476<\beta_{51}<3.467430477$ and $3.373120004<\beta_{53}<3.373120005$. The increasing lower bound $(1+(q-q^{-1})/8)/\log q$ excludes all $q\ge127$ and all $q\ge122$, respectively; the remaining coprime candidates have disjoint score intervals from the winner, with gaps exceeding $0.0159$ and $0.0092$, respectively.

The proof uses divisibility twice: it identifies the exact state at the next pass, and it makes the cumulative squared parent gaps geometrically summable. A general decreasing Shellsort table may change the key sets of the incoming chains at every pass. For $a\ge3$, determining the exact unrestricted leading coefficient remains open. The next section gives pass-budget bounds for general tables and a variational characterization of the unrestricted coefficient when $a=2$. We construct the initial array from independent continuous samples in the source chains. Its expected cost is measured from the resulting ranked array. Charging a prefix that produces this distribution would change the optimization problem.

\section{A pass bound for arbitrary gap tables}\label{sec:vr-arbitrary}

Let $U_{a,m}(p)$ be the minimum mean comparison cost on the same $a$-ordered input when a table may use at most $p$ gaps. The gaps still decrease to $1$, and successive gaps need not divide one another.

\begin{theorem}\label{thm:vr-arbitrary}
For each fixed $a\ge2$, uniformly over integers $p\ge1$ as $m\to\infty$, put $r=\min\{p,\lceil\log(m+1)\rceil\}$. Then
\[
                 U_{a,m}(p)=\Theta_a\bigl(nr\,m^{1/(2r)}\bigr).
\]
The upper bound uses a divisibility table. With no pass limit the optimum over all tables is $\Theta_a(n\log m)$. For each fixed $p$ the exponent of $m$ in the cost per key is $1/(2p)$.
\end{theorem}

Miltersen, Paterson and Tarui use the marginal entropies of the routes of keys to bound the size of a fixed merging network \cite[Section~3.1]{mpt}. Vit\'anyi represents a key's displacement in Shellsort as a signed integer combination of the increments and proves an average lower bound for uniform unrestricted permutations \cite{vitanyi}. Here the input is already ordered on each source chain. We charge each route to the actual comparisons made by the chain insertions. For two chains the joint entropy of the input is only $\log_2\binom{2m}{m}=O(n)$. The sum of the individual rank entropies gives a larger lower bound because the total length of the individual descriptions is controlled by the cost.

\begin{proof}
Use base-two logarithms for entropies. Fix a table of at most $p$ gaps and one realized input. Across a pass of gap $h$, give each key its signed change $t$ in chain index, so its position changes by $ht$. If $M_h$ is the number of successful shifts, then $\sum_{\rm keys}|t|\le 2M_h$. Each shift swaps two keys adjacent on their chain, and each key's net change is bounded in magnitude by its total travel. Write $M=\sum_h M_h$ and $C=\sum_h C_h$. We have $C_h\ge\max(M_h,n-h)$.

Given a key's initial position and the gap table, encode its $t$ at each pass on which its current chain contains at least two keys. Write $|t|$ ones followed by a zero and one extra sign bit if $t\ne0$. The code has length at most $1+2|t|$. The decoder can identify singleton chains from the position computed so far and read no bits there. Since the table length is known, the complete descriptions form a prefix code for the routes from any given start. The final position equals the key's rank. At most $2(n-h)$ keys belong to nonsingleton chains on the $h$-pass, since $s\le2(s-1)$ for each chain length $s\ge2$. The total code length for all keys on this input is at most
\[
              \sum_h[2(n-h)+4M_h]\le6C.
\]

Select one of the original independent continuous samples from chain $c$. Conditional on its value $u$ and the count $k$ of smaller samples in its chain, this marked key starts at $ak+c$. Its rank is $k+Y$ with $Y\sim{\rm Bin}(N,u)$, where $N=(a-1)m$. The other chains are independent of the conditioning. The conditional entropy of its rank, averaged over $u,k$, is at least
\[
                 \tfrac12\log_2 N-O(1)\ \ge\ \tfrac12\log_2 n-O(1).
\]
Indeed, the largest atom of the binomial law is at most $K/\sqrt{Nu(1-u)}$ for an absolute $K$. Fourier inversion bounds each atom by the integral of $|1-u+u e^{2\pi i x}|^N$ on $[-1/2,1/2]$. Using $\sin(\pi |x|)\ge2|x|$ bounds this integral by that of $\exp(-8Nu(1-u)x^2)$ over the real line. Entropy is at least minus the logarithm of the largest atom, and the integral of $\ln(u(1-u))$ on $(0,1)$ is $-2$. Apply the prefix-code bound for each mark with its start known, and sum over all $n$ original samples. This proves $\mathbb E C=\Omega(n\log n)$ for every table, with an absolute constant.

To retain the dependence on $p$, give the same route as a vector of $p$ signed integers, padding with zeros after the last pass. For any integer-valued random variable $T$ with $\mathbb E|T|=v$, its entropy is at most $\log_2(2e(1+v))$. Its sign costs at most one bit, and the geometric law of mean $v$ maximizes entropy on the nonnegative integers of that mean. The bound follows by nonnegativity of relative entropy against that law and $(1+1/v)^v\le e$. The case $v=0$ is immediate. Given the start, the vector determines the rank. Apply entropy subadditivity conditional on $u,k$, then use concavity of the logarithm across the conditioning variables, the $p$ coordinates and the $n$ marks. Using $\sum_{\rm keys}\sum_h |t|\le2M$ gives
\[
  \tfrac12\log_2 n-O(1)
          \le p\log_2\!\left(2e\left(1+\frac{2\mathbb E M}{np}\right)\right).
\]
If $p=r$ and $x=m^{1/(2p)}$ exceeds a sufficiently large absolute constant, exponentiation gives $\mathbb E C\ge\mathbb E M=\Omega(npx)$. The factor arising from $O(1)/p$ is bounded uniformly in $p$. If $p=r$ and $x$ is bounded, $npx=O(n\log m)$, already covered by the route-code bound. For $p>r$, $r=\Theta(\log m)$ and $m^{1/(2r)}=O(1)$, so the same bound applies.

For the upper bound take the least integer $q\ge m^{1/(2r)}$ congruent to $1\bmod a$. For $m\ge2$, $2\le q\le m^{1/(2r)}+a$. Put
\[
        j=\min\{r-1,\lfloor\log_q\sqrt m\rfloor\},\qquad H=(q^j,q^{j-1},\ldots,1).
\]
For $j=0$ take the single gap $1$. The first gap $g=q^j$ lies between $\sqrt m/q$ and $\sqrt m$, because $q^r\ge\sqrt m$. Equations~\eqref{eq:vr-foot} and \eqref{eq:vr-entry} give $\mathbb E C_g=O_a(nq)$. Before a later step $L\to h=L/q$, the array is ordered at gaps $a$ and $L$ by preservation of earlier orders. They are coprime, so every positive distance at least $(a-1)(L-1)$ is ordered, as shown before \eqref{eq:vr-cross}. Each key has at most $(a-1)L/h=(a-1)q$ inverted predecessors in its $h$-chain. These passes each cost $O_a(nq)$, and there are at most $r$ passes in total.
\end{proof}

\begin{corollary}\label{cor:vr-free-coefficient}
Let $\tau>1$ solve $\tau^3=\tau^2+\tau+1$. For each fixed $a\ge2$, let $U_{a,m}$ be the minimum over all decreasing tables, with no divisibility or pass constraint. Then
\[
 \frac{n\log m}{4\log\tau}-O_a(n)\ \le\ U_{a,m}\
 \le\ \frac{\beta_a}{2}\,n\log m+O_a(n).
\]
For two source chains the respective leading coefficients are $1/(4\log\tau)$ and $13/(14\log 7)$, approximately $0.4103$ and $0.4772$. Theorem~\ref{thm:vr-separation} improves the upper endpoint in this case.
\end{corollary}
\begin{proof}
Take a minimizing table and omit its initial gaps $h\ge n^{3/4}$. The far-inversion estimate in Theorem~\ref{thm:vr-first-gap} bounds the probability of any initial inversion at such a distance by $2n^2\exp(-\sqrt n/(2a))$. On its complement all omitted passes leave the input fixed. On the exceptional event even the whole shortened table costs $O(n^3)$ comparisons. Thus its mean cost $\bar C$ is at most $U_{a,m}+o(1)$ and is $O_a(n\log m)$ by Theorem~\ref{thm:vr-main}.

Write $p$ for the remaining pass count, $s=\sum_h h$ for their total number of chains, and $R$ for the total number of noninitial record minima in them. Each gap is eventually below $n/2$, so $p=O_a(\log m)$ and $s=O_a(n^{3/4}\log m)=o(n)$. If a chain has $b$ new record minima, these keys incur at least $1+2+\cdots+b$ shifts in total. Cauchy--Schwarz over the chains gives $R\le\sqrt{2sM}$ and hence $\mathbb E R=O_a(n^{7/8}\log m)=o(n)$ since $\mathbb E M\le\bar C$. The exact comparison identity \eqref{eq:vr-count} now gives
\[
                    np+\mathbb E M=\bar C+s+\mathbb E R
                                    =\bar C+o(n).
\]

For the displacement vector in Theorem~\ref{thm:vr-arbitrary}, bound entropy in natural units by cross entropy against the product of the integer laws $\Pr(t)=\tau^{-|t|}/Z$. Here $Z=(\tau+1)/(\tau-1)=\tau^2$. The conditional rank entropy summed over all marks is at least $(n/2)\log m-O(n)$ by the binomial estimate. The sum of the absolute net displacements over all marks and passes is at most $2M$. Consequently
\[
       \tfrac n2\log m-O(n)\le np\log Z+2(\log\tau)\mathbb E M
              =2(\log\tau)\,[\bar C+o(n)] .
\]
The upper bound uses the divisibility table of Theorem~\ref{thm:vr-main}.
\end{proof}

\begin{theorem}\label{thm:vr-separation}
For $a=2$ there is a decreasing table with mean comparison cost
\[
             \frac{86}{47\log 47}\,n\log m+O(n).
\]
Its leading coefficient is strictly below $13/(14\log 7)$, the minimum for divisibility tables. In particular $U_{2,m}\le [86/(47\log 47)]n\log m+O(n)$.
\end{theorem}
\begin{proof}
Start with $47^J$ for $J=\lfloor\log_{47}\sqrt m\rfloor$. After each anchor $L=47^j$ with $j\ge1$, run the two gaps $7t,t$ for $t=L/47$. Thus the last two blocks use $2209,329,47,7,1$, counting the initial anchor of those blocks once. The first pass costs $O(n)$ in expectation by \eqref{eq:vr-foot} and \eqref{eq:vr-entry}.

For a marked key write $D$ for its current physical position minus its final rank. On any $2$-ordered array all of its inversion partners lie on the other parity and on a single side. Indeed a larger predecessor and a smaller successor would contradict the ordering of the opposite parity. Thus $|D|$ counts its inversion partners. After an odd $h$-pass all odd distances at least $h$ are ordered, since they have the form $h+2k$ and the $2$-order is preserved. The key has at most $(h-1)/2$ inversion partners, and its displacement is congruent to its previous value modulo $h$. Consequently its new displacement is the centered residue $\operatorname{rem}_h(D)$ in $[-(h-1)/2,(h-1)/2]$. In the pass the key crosses only inversion partners on that one side. Its absolute net motion in chain coordinates therefore counts all its crossings. Summing these absolute motions over the keys counts every successful shift twice.

All gaps before an anchor $L$ are multiples of $L$. Immediately after it the marked displacement is therefore the centered residue of its initial displacement modulo $L$. Write it uniquely as $tk+e$, where $-23\le k\le23$ and $|e|\le(t-1)/2$. Set $b=\lfloor(k+3)/7\rfloor$, the integer nearest to $k/7$. The nearest integer to $(tk+e)/(7t)$ is also $b$. The two successive gaps $7t,t$ thus move this key by $|b|$ and $|k-7b|$ chain positions, respectively, and leave centered residue $e$. If the initial displacement modulo $L$ were uniform, the $47$ values of $k$ would be equiprobable. By symmetry, the mean contribution to the two-pass shift count per key (including the factor $1/2$) would be
\[
   \frac1{47}\sum_{k=1}^{23}
       \left(\left\lfloor\frac{k+3}{7}\right\rfloor+
       \left|k-7\left\lfloor\frac{k+3}{7}\right\rfloor\right|\right)
       =\frac{78}{47}.
\]

For an original sample on parity $c$, conditional on its value $u$ and the number $k_0$ of smaller samples on that parity, its initial displacement is $c+k_0-Y$, $Y\sim{\rm Bin}(m,u)$. The total variation estimate in Lemma~\ref{lem:vr-pass} bounds the distance of this residue from uniform, integrated over $u$, by $\pi^2 L^2/(12m)$. Translation and reflection modulo $L$ preserve distance from uniform. The shift contribution of a key is bounded by an absolute constant, so the error in the block expectation is $O(nL^2/m)$. Summed over $L=47^J,\ldots,47$, these errors are $O(n)$.

On any $h$-chain, only its second key can set a new record minimum after its first key: each later key exceeds the key two slots earlier by $2$-order. Thus $R_h\le h$ and the exact count \eqref{eq:vr-count} gives $C_h=n+M_h+O(h)$. The sum of the gaps in the blocks is $O(\sqrt m)$. The total mean cost is consequently $nJ(2+78/47)+O(n)$, as asserted. Finally $86/(47\log 47)<13/(14\log 7)$ is equivalent to $18\log 7>611\log(49/47)$. The left side exceeds $27$ and the right side is less than $611\cdot2/47=26$.
\end{proof}

\begin{corollary}\label{cor:vr-three-step}
For two starting chains there is an unrestricted decreasing table with mean comparison count
\[
  \frac{1094}{387\log 387}\,n\log m+O(n).
\]
\end{corollary}
\begin{proof}
Use anchors \(387^J,\ldots,387,1\), with \(J=\lfloor\log_{387}\sqrt m\rfloor\). From each anchor \(387t\) run \(59t,7t,t\). At the anchor write the centered displacement as \(tk+e\), with \(-193\le k\le193\) and \(|e|\le(t-1)/2\). For each \(s=59,7,1\) in turn, let \(b_s\) be the nearest integer to \(k/s\) and replace \(k\) by \(k-sb_s\). This gives exactly the chain motion of the marked key, because \(|t(k-sb_s)+e|\le(ts-1)/2\) before replacing \(k\). Starting with \(k=1,\ldots,193\), the sums of \(|b_s|\) are respectively \(315,391,321\); these integers follow by using \(b_s=\lfloor(k+(s-1)/2)/s\rfloor\) at each step. The terms for negative starting \(k\) are equal by symmetry. Thus the reference mean number of shifts per key across the block is \(1027/387\), including the factor \(1/2\) for counting both keys crossed by a shift.

As in Theorem~\ref{thm:vr-separation}, all gaps before an anchor are its multiples, so the displacement there is the centered residue of the initial displacement. The bounded three-pass motion has mean error \(O(n(387t)^2/m)\) per block from replacing this residue by a uniform one. The errors sum to \(O(n)\), the first anchor costs \(O(n)\), and the gap sum is \(O(\sqrt m)\). The three scan charges add \(3n+O(387t)\) per block. The total is \(nJ(3+1027/387)+O(n)\), giving the displayed coefficient. It is strictly smaller than \(86/(47\log47)\), equivalently \(47^{25709}<387^{16641}\).
\end{proof}

\subsection{The leading coefficient for all two-chain tables}

An odd block consists of an odd integer \(B\ge3\) and a list \(H=(h_1,\ldots,h_p)\) of decreasing odd gaps below \(B\), ending at \(1\). For an integer \(d_0=d\), let \(d_i=\operatorname{rem}_{h_i}(d_{i-1})\) be the centered residue and define
\[
 f_H(d)=\frac12\sum_{i=1}^p\left|\frac{d_{i-1}-d_i}{h_i}\right|,\qquad
 \mu(B,H)=\frac1B\sum_{|d|\le(B-1)/2} f_H(d).
\]
Thus \(f_H\) charges to one marked key half its chain motion. Put
\[
 \gamma=\inf_{(B,H)}\frac{p+\mu(B,H)}{2\log B}.
\]

\begin{theorem}\label{thm:vr-two-variation}
For a uniform interleaving of two sorted chains,
\[
   \lim_{m\to\infty}\frac{U_{2,m}}{n\log m}=\gamma,\qquad
   \frac1{4\log\tau}\le\gamma\le\frac{1094}{387\log 387}.
\]
Each fixed block supplies its stated coefficient with an additive \(O_{B,H}(n)\) error. The lower bound \(U_{2,m}\ge\gamma n\log m-O(n)\) holds for the unrestricted optimum.
\end{theorem}
\begin{proof}
For the upper bound, repeat \(H\) below anchors \(B^J,\ldots,B,1\), where \(J=\lfloor\log_B\sqrt m\rfloor\). Below \(Bt\) execute \(th_1,\ldots,th_p=t\). The first anchor costs \(O_B(n)\) in the mean by \eqref{eq:vr-foot} and \eqref{eq:vr-entry}. At any anchor \(Bt\) the displacement is the centered residue of its initial value there, since earlier gaps were multiples of \(Bt\). Write this residue as \(tk+e\), with \(|k|\le(B-1)/2\) and \(|e|\le(t-1)/2\). For odd \(h\), \(\operatorname{rem}_{th}(tk+e)=t\operatorname{rem}_h(k)+e\). By the displacement identity in Theorem~\ref{thm:vr-separation}, this block charges \(f_H(k)\) shifts per marked key, including its half share of each crossing. A uniform residue modulo \(Bt\) makes \(k\) uniform, and the error for the actual residue is \(O_{B,H}((Bt)^2/m)\) in the mean per key, by the integrated total variation bound in Lemma~\ref{lem:vr-pass} applied to the opposite chain. The errors sum to \(O_{B,H}(n)\) over all keys and blocks. Each pass costs \(n+M_h+O(h)\) by two-chain order, and the gap sum is \(O_{B,H}(\sqrt m)\). Hence the cost is \(nJ[p+\mu(B,H)]+O_{B,H}(n)\). Taking the infimum after the limit superior gives the upper estimate.

For the converse, mark one original key uniformly at random and let \(D\) be its initial position minus its rank. Its law \(\pi\) is symmetric and nonincreasing on nonnegative integers, with \(\pi(0)=O(m^{-1/2})\). To prove this, condition on the continuous value \(u\) of the mark. Its source parity \(c\) is uniform on \(\{0,1\}\). The counts of smaller values in its own and opposite chains give
\[
 D=c+K-Y'-e,\qquad K,Y'\ \text{independent }{\rm Bin}(m-1,u),
 \qquad e\sim{\rm Bernoulli}(u),
\]
where \(e,c,K,Y'\) are conditionally independent. Set \(v=u(1-u)\). The law of \(K-Y'\) is the convolution of \(m-1\) copies of the law \((v,1-2v,v)\) on \(-1,0,1\). Pairing \(u\) with \(1-u\) preserves this convolution and makes the independent offset \(c-e\) have law \((1/4,1/2,1/4)\). Convolution with \((v,1-2v,v)\), \(0\le v\le1/4\), preserves symmetric unimodality: it is a convex combination of a point mass at zero and the uniform law on \(\{-1,0,1\}\), and a centered three-term sliding average of a symmetric unimodal sequence has the same property. For the central mass, Fourier inversion bounds the largest atom of \(K-Y'\) by
\[
  \frac1{2\pi}\int_{-\pi}^{\pi}
       (1-4v\sin^2(\theta/2))^{m-1}\,d\theta
       =O\bigl(((m-1)v)^{-1/2}\bigr)
\]
for \(m\ge2\), using \(\sin(|\theta|/2)\ge|\theta|/\pi\). Adding an independent offset preserves this upper bound. Integrating in \(u\) uses \(\int_0^1du/\sqrt{u(1-u)}=\pi\), proving the claim.

The weights \(\lambda_r=(2r+1)(\pi(r)-\pi(r+1))\), \(r\ge0\), now express \(\pi\) as a probability mixture of uniform laws on \([-r,r]\). Indeed \(\sum_{r\ge|d|}\lambda_r/(2r+1)=\pi(d)\). For the random length \(B'=2r+1\) under these weights, \(\Pr(B'\le x)\le x\pi(0)\). Integrating this bound at \(x=\sqrt m e^{-t}\) for \(t\ge0\), truncated at one, yields \(\mathbb E\log B'\ge\tfrac12\log m-O(1)\).

Even gaps make no moves on a two-chain input and may be dropped. Remove the initial gaps at least \(n^{3/4}\) from an optimal table as in the proof of Corollary~\ref{cor:vr-free-coefficient}. The shortened odd table \(G\) has \(p'\) passes and satisfies \(np'+\mathbb E M\le U_{2,m}+o(n)\). Its mean shifts obey \(\mathbb E M/n=\mathbb E f_G(D)\) by summing the marked-key identity. On an interval of odd length \(B'\ge3\) in the mixture, every gap at least \(B'\) leaves every displacement fixed. Removing these gaps gives a legal block at \(B'\) with the same \(f_G\) on the interval. Thus \(p'+\mathbb E[f_G(D)\mid B']\ge2\gamma\log B'\), also valid for \(B'=1\). Averaging proves \(U_{2,m}\ge\gamma n\log m-O(n)\). The explicit endpoints follow from Corollaries~\ref{cor:vr-free-coefficient} and~\ref{cor:vr-three-step}.
\end{proof}
The formula characterizes the coefficient by finite blocks of unbounded size. Its exact value and the existence of a block attaining the infimum remain open.

\appendix
\section{Exact optimization at finite length}\label{sec:vr-finite}

The asymptotic construction can be compared with an exact optimum for every \(a,m\). Read the original chain labels in increasing order of key rank. The resulting word is uniform over the \(W=\binom{n}{m,\ldots,m}\) words with \(m\) copies of each label. Write \(\binom{|\boldsymbol v|}{\boldsymbol v}=|\boldsymbol v|!/\prod_d v_d!\), where \(|\boldsymbol v|=\sum_d v_d\). For a prefix with label counts \(\boldsymbol k=(k_0,\ldots,k_{a-1})\), the probability that these counts occur and that the next letter is \(c\) is
\[
 w(\boldsymbol k,c)
 =W^{-1}\binom{|\boldsymbol k|}{\boldsymbol k}
   \binom{n-|\boldsymbol k|-1}{m\boldsymbol 1-\boldsymbol k-\boldsymbol e_c}.
\]
Here \(0\le k_d\le m\) and \(k_c<m\). The two coefficients count orders before and after the marked letter.

Let \(L\to h\) denote a pass at gap \(h\) starting from \(S_LA\). Introduce a virtual source \(L=n\), for which \(S_nA=A\). The entrance positions of keys below the mark form \(E=\{aj+d:0\le d<a,\ 0\le j<k_d\}\), and the mark starts at \(i=ak_c+c\). Set \(N_s=\#\{j\in E:j\equiv s\pmod L\}\). After \(S_L\) the set of smaller positions and the marked position are respectively
\[
 E_L=\{s+Lj:0\le s<L,\ 0\le j<N_s\},\qquad
 p=(i\bmod L)+LN_{i\bmod L}.
\]
Among the \(\lfloor p/h\rfloor\) earlier positions of its \(h\)-chain, let \(b\) count the ones in \(E_L\). When this mark is inserted, the earlier keys have already been sorted within the chain. Each of its larger predecessors costs one successful comparison, and the presence of a smaller predecessor costs one terminating comparison. Summing over marks therefore gives the exact edge weight
\[
 c(L,h)=\sum_{\boldsymbol k,c} w(\boldsymbol k,c)
 \left(\lfloor p/h\rfloor-b+\mathbf1_{\{b>0\}}\right).
\]
For each rank the corresponding weights sum to one, and the displayed sum includes all \(n\) ranks.

Use vertices \(n,n-1,\ldots,1\), with an edge from \(n\) to every smaller gap and from each other \(L>1\) to every proper divisor \(h\) of \(L\). Each path to \(1\) gives exactly one legal table. Its expected cost is the sum of its edge weights because the actual incoming state on each internal edge is \(S_L A\). Thus a shortest path constructs an exact optimal table with rational cost; computing the weights requires at most \(a(m+1)^a\) choices of the marked prefix data per edge. The following instances were independently checked by direct insertion sorting on every rank word and every legal table. The listed optimum is unique in each row.
\[
 \begin{array}{c|r|r|c|c}
 (a,m)&\text{rank words}&\text{tables}&\text{optimal table}&\min_H{\mathbb E}C(H;A)\\ \hline
 (2,3)&20&6&(1)&69/10\\
 (2,4)&70&10&(1)&711/70\\
 (2,5)&252&16&(1)&1711/126
 \end{array}
\]

At \(a=2,m=75\), the shortest-path computation covers 1,848 legal tables and
\(\binom{150}{75}\) equally likely rank words. The unique minimizer is
\((5,1)\), with expected cost \(436.37160943\) to eight decimal places.
The runner-up is \((1)\), with expected cost \(436.79071740\).
All comparisons of table costs use integer numerators over the common
denominator \(\binom{150}{75}\). The two reported costs can also be
verified by a forward recursion on the counts of the two letters in a
rank-word prefix. Appending a letter adds exactly one low position after
\(S_L\). Charge to its key one inversion for each earlier-ranked key to its
right on its \(h\)-chain, and subtract one record when its preceding
chain segment is nonempty and contains no earlier-ranked key. After summing
these charges and adding \(n-h\) for the scan, this recursion gives the same
edge totals for both tables. The finite minimizer illustrates why the
optimal asymptotic quotient \(7\) need not appear in a short table.

\section*{Use of generative AI}
ChatGPT/Codex assisted with English writing, translation, typesetting, editorial checking of mathematical statements, literature and submission-requirement checks, and implementation and execution of finite verification checks. This revision is based on the mathematical arguments in the source manuscript supplied by the author. The author retains responsibility for the content of this manuscript. The finite checks accompanying the paper verify selected identities and instances; they do not replace its mathematical proofs.

\end{document}